\documentclass[nocompress]{spie}  

\usepackage{amsmath,amsfonts,amssymb}
\usepackage{graphicx}
\usepackage[colorlinks=true, allcolors=blue]{hyperref}
\usepackage{float}
\usepackage{subfig}

\title{SPHERExLabTools (SLT): A Python Data Acquisition System for SPHEREx Characterization and Calibration}

\author[a]{Sam Condon}
\author[a]{Marco Viero}
\author[a, b]{James Bock}
\author[a]{Howard Hui}
\author[a]{Phil Korngut}
\author[a]{Hiromasa Miyasaka}
\author[b]{Ken Manatt}
\author[a]{Chi Nguyen}
\author[b]{Hien Nguyen}
\author[a]{Steve Padin}
\affil[a]{California Institute of Technology, Pasadena, CA, 91125, USA}
\affil[b]{Jet Propulsion Laboratory, California Institute of Technology, Pasadena, CA, 91109, USA}
\authorinfo{Send correspondence to S. Condon: scondon@caltech.edu}

\begin{document}
\maketitle
 
\begin{abstract}
Selected as the next NASA Medium Class Explorer mission, SPHEREx, the Spectro-Photometer for the History of the Universe, Epoch of Reionization, and Ices Explorer is planned for launch in early 2025. SPHEREx calibration data products include detector spectral response, non-linearity, persistence, and telescope focus error measurements. To produce these calibration products, we have developed a dedicated data acquisition and instrument control system, SPHERExLabTools (SLT). SLT implements driver-level software for control of all testbed instrumentation, graphical interfaces for control of instruments and automated measurements, real-time data visualization, processing, and data archival tools for a variety of output file formats. This work outlines the architecture of the SLT software as a framework for general purpose laboratory data acquisition and instrument control. Initial SPHEREx calibration products acquired while using SLT are also presented.
\end{abstract}

\keywords{SPHEREx, All-sky survey, ground-support-equipment, data-acquisition}

\section{INTRODUCTION}
\label{sec:intro}  

The SPHEREx mission will perform an all-sky spectrophotometric survey in the near-infrared (0.75 $\mu$m. - 5.0 $\mu$m.) to address major scientific questions consistent with all three major themes of NASA's astrophysics program. Specifically, SPHEREx will:

\begin{enumerate}
    \item \textit{Probe the origin and destiny of our Universe.} By measuring galaxy redshifts over a large cosmological volume, SPHEREx provides constraints on the physics of inflation, the superluminal expansion of the Universe that took place $10^{-32}$ s after the Big Bang. 
    \item \textit{Explore whether planets around other stars could harbor life.}  SPHEREx provides absorption spectra of early phase planetary systems in the Milky Way to determine the abundance and composition of biogenic ices.
    \item \textit{Explore the origin and evolution of galaxies.} SPHEREx will permit a precise determination of the power spectrum of spatial fluctuations in the extragalactic background light intensity through deep images formed near the ecliptic poles.
\end{enumerate}

SPHEREx instrument parameters are listed in table \ref{table:InstrumentParams}. Previous SPIE proceedings \cite{Korngut18} and \cite{Crill20} provide detailed information about the scientific rationale, hardware configuration, and mission design of SPHEREx. 

By mid-2022, SPHEREx is in its optical payload integration and testing phase to deliver several laboratory calibration and instrument characterization measurements. Two such measurements are discussed in this paper: absolute spectral and focus calibration. To accurately determine photometric redshifts and resolve ice absorption features in the spectra of planetary systems, precise spectral bandpass profiles for all 25 million pixels in SPHEREx must be determined through spectral calibration. To maximize the fraction of background sky pixels and enable source masking to deeper flux levels for extragalactic background light power spectrum measurements, the SPHEREx point-spread-function (PSF) must be minimized through focus calibration. A system capable of performing these characterization/calibration measurements must meet the requirements listed in Table \ref{table:ControlReqs}.

\begin{table}[h]
\caption{SPHEREx Instrument Parameters} 
\label{table:InstrumentParams}
\begin{center}       
\begin{tabular}{|l|l|}
\hline
\rule[-1ex]{0pt}{3.5ex}  \textbf{Parameter} & \textbf{Value}  \\
\hline
\rule[-1ex]{0pt}{3.5ex}  Effective Aperture & 20 cm \\
\hline
\rule[-1ex]{0pt}{3.5ex}  Pixel Size & 6.2" x 6.2" \\
\hline
\rule[-1ex]{0pt}{3.5ex}  Field of View & 2 x (3.5\textdegree x 11.3\textdegree); dichroic\\
\hline
\rule[-1ex]{0pt}{3.5ex}  Resolving Power and Wavelength Coverage & 
    \begin{tabular}{l}
      $\lambda = 0.75 - 2.42 \mu$m; $R=41$ \\
      $\lambda = 2.42 - 3.82 \mu$m; $R=35$ \\
      $\lambda = 3.82 - 4.42 \mu$m; $R=110$ \\
      $\lambda = 4.42 - 5.00 \mu$m; $R=130$ \\
    \end{tabular} \\
\hline
\rule[-1ex]{0pt}{3.5ex}  Arrays & 
    \begin{tabular}{l}
      3x Teledyne Hawaii-2RG 2.5 $\mu$m. \\
      3x Teledyne Hawaii-2RG 5.3 $\mu$m. \\
    \end{tabular} \\
\hline
\rule[-1ex]{0pt}{3.5ex}  Cooling & All-Passive \\
\hline
\rule[-1ex]{0pt}{3.5ex}  2.5 $\mu$m. Array Temperature & $<80K$ \\
\hline
\rule[-1ex]{0pt}{3.5ex}  5.3 $\mu$m. Array Temperature & $<55K$ \\
\hline
\end{tabular}
\end{center}
\end{table} 

\begin{table}[h]
\caption{SPHEREx characterization/calibration system requirements.} 
\label{table:ControlReqs}
\begin{center}       
\begin{tabular}{|l|l|} 
\hline
\rule[-1ex]{0pt}{3.5ex}  \textbf{Requirement} & \textbf{Description}  \\
\hline
\rule[-1ex]{0pt}{3.5ex} 1) Instrument driver framework & \begin{tabular}{l}
      A convenient abstraction away from low-level instrument \\
      communication details.
    \end{tabular} \\
\hline
\rule[-1ex]{0pt}{3.5ex} 2) Manual instrument control & 
    \begin{tabular}{l}
     Interfaces to control instruments in a \\
     testbed to facilitate the setup of automated \\
     measurements.
    \end{tabular} \\
\hline
\rule[-1ex]{0pt}{3.5ex} 3) Measurement automation & 
    \begin{tabular}{l}
     A script based framework for specifying automated \\
     measurements.
    \end{tabular} \\
\hline
\rule[-1ex]{0pt}{3.5ex} 4) Real-time data visualization and processing & \begin{tabular}{l}
        Live plots of multi-channel time-stream data and image \\
        feeds. Basic real-time processing such as averaging \\
        and histogram display.
    \end{tabular} \\
\hline
\rule[-1ex]{0pt}{3.5ex} 5) Data archival tools & \begin{tabular}{l}
        Mechanisms to archive data products along with \\
        associated metadata in a variety of file formats.
    \end{tabular} \\
\hline
\rule[-1ex]{0pt}{3.5ex} 6) Configurability & 
    \begin{tabular}{l}
     A method to reconfigure all aforementioned \\
     components for a variety of measurements. 
    \end{tabular} \\
\hline
\end{tabular}
\end{center}
\end{table} 

The development of a software system meeting the Table \ref{table:ControlReqs} requirements is a problem that is not unique to SPHEREx. Modern experimental laboratories everywhere are faced with the similar challenge of developing such a system. Despite this, there are limited options when it comes to standard, well tested data acquisition and instrument control packages providing these capabilities. Familiar to many is NI's LabView \cite{LabView}. LabView has the advantage of being well tested, widely adopted, and with the ample support expected from any large-scale proprietary software system. However, integrating custom hardware with LabView can be cumbersome and the graphical programming environment lacks the flexibility that a true general purpose programming language provides. In the open source landscape, the PyMeasure project \cite{PyMeasure} provides a Python based measurement automation package. PyMeasure contains a large repository of existing instrument drivers along with a convenient framework for adding new instruments. Measurement automation tasks are easily achieved and measurement control logic is specified through basic Python scripts. In addition, PyMeasure has gathered a very active development community to support further improvements to the package. However, \emph{.csv} is currently the only output file format supported and the package does not have a convenient mechanism for manual instrument control. PyHK \cite{PyHk} is an additional Python based open source data acquisition package with adoption in astrophysics and astronomy instrumentation laboratories such as the BICEP \cite{BICEP} and TIME \cite{TIME} collaborations. Initially developed at Caltech for control and monitoring of cryogenic instruments, PyHK provides an excellent web-based user interface for manually controlling instruments, as well as viewing and interacting with housekeeping data stored over long periods of time. PyHK also contains a robust configuration file based interface for specifying the instruments in use and to customize a web interface for a given experiment. PyHK's limitations come from the fact that the instrument driver framework is not as robust as PyMeasure's and plain text is the only supported output file format. 

Following the evaluation outlined above, it was determined that a custom data acquisition and instrument control system would best suit the needs of SPHEREx characterization/calibration. \textbf{SPHERExLabTools}, hereafter referred to as SLT, is the resulting system. SLT meets all of the Table \ref{table:ControlReqs} requirements and presents a framework for a general purpose laboratory data acquisition and instrument control system. SLT adopts the same instrument driver framework and method of specifying automated measurements as PyMeasure while providing a configuration file based interface that is similar to PyHK. In addition, SLT addresses the limitations of PyMeasure and PyHK by providing support for multiple output file formats, while integrating both manual and automated measurement control mechanisms.

The structure of the remainder of this paper is as follows. Section \ref{section:SLT Architecture} describes the architecture and configuration mechanism of SLT, sections \ref{section:Spectral Cal} and \ref{section:Focus Cal} describe the SPHEREx spectral and focus calibration measurements and illustrate how SLT is configured for each measurement. Initial calibration results obtained using SLT are also presented in these sections. Finally, section \ref{section:Conclusion} provides concluding remarks and discusses plans for further development of the SLT system.

\section{SLT Architecture} \label{section:SLT Architecture}
 
\subsection{Implementation Language and Class Structure} 
 
\begin{figure}[t]
    \centering
    \includegraphics[width=0.8\linewidth]{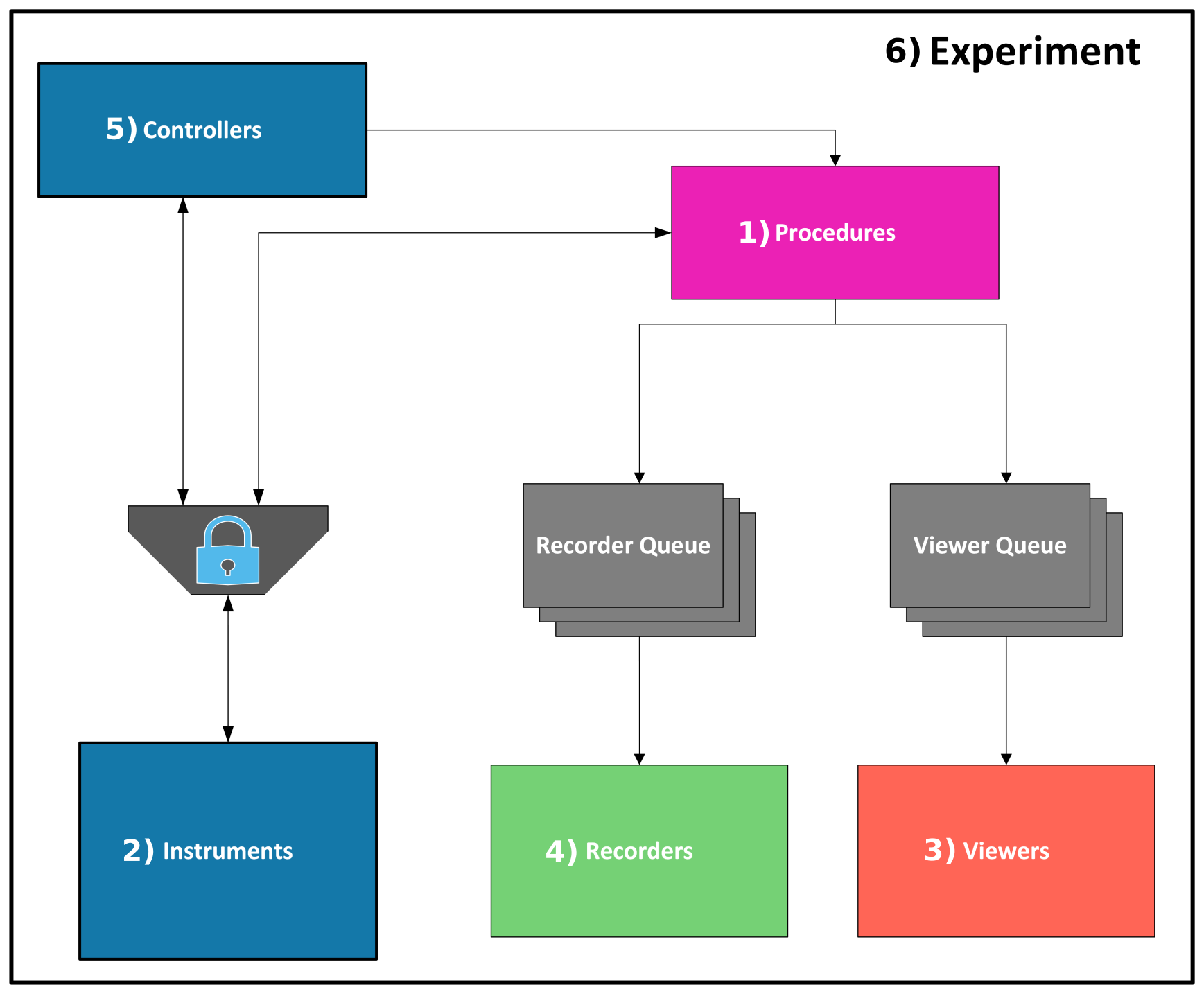}
    \caption{The SLT modular architecture consists of 6 classes which interact in the following manner: \textbf{1) Procedures} contain scripts which define the logical flow of a measurement; they interact directly with \textbf{2) Instruments} and generate data which they send to \textbf{3) Viewers} for real time graphical display and/or \textbf{4) Recorders} to archive data. Recorders and Viewers sit idle until data is placed on their associated queue, upon which they perform the appropriate archival or display task. \textbf{5) Controllers} provide graphical interfaces to interact directly with Instruments or to start/stop Procedure execution. Since both Controllers and Procedures have the ability to interact directly with Instruments, a locking mechanism is placed on Instruments such that only one controller/procedure thread of execution can access any given instrument driver class at a single instant. Finally, the \textbf{6) Experiment} class provides a top-level wrapper to interpret user configuration files and create instances of all of the previous classes. SLT is a multi-threaded application so many instances of each of these components can execute simultaneously. The color-scheme in this figure indicates the threading structure where each color corresponds to a separate thread of execution.}
    \label{fig:SLT_Architecture}
\end{figure}

 SLT is implemented in pure Python. This decision was motivated by the recent wide adoption of Python in many scientific communities and the existence of the powerful Python data ecosystem including packages like NumPy \cite{harris2020array}, SciPy \cite{SciPy}, and Pandas \cite{mckinney-proc-scipy-2010}. The implementation of SLT in Python allows us to utilize this ecosystem while working in an environment familiar to many scientists involved in SPHEREx and beyond. In addition, Python's object-oriented nature facilitates the development of organized modular code for which each requirement of Table \ref{table:ControlReqs} can be met through independent configurable modules and/or classes. The SLT class based modular architecture is summarized in Figure \ref{fig:SLT_Architecture} while section \ref{sect:ClassDescription_UI} describes in more detail all of the classes outlined in this figure.
 
\subsection{Experiment Control Packages and Configurability}

SLT focuses heavily on configurability to support the diversity of measurements and instruments involved in SPHEREx testing. The manner in which SLT is configured is through the specification of \textbf{\emph{Experiment Control Packages}}. Experiment control packages are Python packages that define the 5 configuration variables described in Table \ref{table:ConfigVariables}. In brief, these configuration variables are read by the SLT Experiment class (6) of Figure \ref{fig:SLT_Architecture}) to command SLT which instrument drivers and graphical interfaces to instantiate, what the output data formats should be, and where to find user-defined measurement automation scripts (i.e. Procedures, sect. \ref{sect:Procedures}). Table \ref{table:ConfigVariables} describes the purpose of each of these configuration variables, while Tables \ref{table:ProcConfig} through \ref{table:CntrlConfig} specify each variable's appropriate contents. 

\begin{table}[H]
\caption{SLT Experiment control package configuration variables. Each variable is a python list of \textbf{configuration dictionaries} (see tables \ref{table:ProcConfig} - \ref{table:CntrlConfig}). The numbered elements of Figure \ref{fig:SLT_Architecture} correspond to the variable numbers in this table.} 
\label{table:ConfigVariables}
\begin{center}       
\begin{tabular}{|l|l|} 
\hline
\rule[-1ex]{0pt}{3.5ex}  \textbf{Variable} & \textbf{Description}  \\
\hline
\rule[-1ex]{0pt}{3.5ex} 1) PROCEDURES & 
\begin{tabular}{l}
	List of \textbf{procedure configuration dictionaries} (Table \ref{table:ProcConfig}) pointing SLT to the \\
	user defined measurement automation scripts i.e. (Procedure classes). 
\end{tabular} \\
\hline
\rule[-1ex]{0pt}{3.5ex} 2) INSTRUMENT\_SUITE & 
    \begin{tabular}{l}
    List of \textbf{hardware configuration dictionaries} (Table \ref{table:HardwareConfig}) telling SLT \\
    which instrument drivers should be instantiated. 
    \end{tabular} \\
\hline
\rule[-1ex]{0pt}{3.5ex} 3) VIEWERS & 
\begin{tabular}{l}
	List of \textbf{viewer configuration dictionaries} (Table \ref{table:ViewConfig}) defining data visualization \\
	tasks.
\end{tabular} \\
\hline
\rule[-1ex]{0pt}{3.5ex} 4) RECORDERS & 
\begin{tabular}{l}
	List of \textbf{recorder configuration dictionaries} (Table \ref{table:RecConfig}) to specify the output file \\
	formats that data generated from procedures should be saved to.
\end{tabular} \\
\hline
\rule[-1ex]{0pt}{3.5ex} 5) CONTROLLERS & 
    \begin{tabular}{l}
    List of \textbf{controller configuration dictionaries} (Table \ref{table:CntrlConfig}) identifying the \\
    instruments and procedures for which graphical control interfaces \\
    should be generated.
    \end{tabular} \\
\hline
\end{tabular}
\end{center}
\end{table} 

\begin{table}[H]
	\caption{SLT \textbf{procedure configuration dictionary} format. The \textbf{1) PROCEDURE} variable is a python list of dictionaries with these key/value pairs.} 
	\label{table:ProcConfig}
	\begin{center}       
		\begin{tabular}{|l|l|} 
			\hline
			\rule[-1ex]{0pt}{3.5ex}  \textbf{Key} & \textbf{Value Description}  \\
			\hline
			
			\rule[-1ex]{0pt}{3.5ex} instance\_name & 
			\begin{tabular}{l}
				String to provide a name for the procedure instance.
			\end{tabular} \\
			\hline
			
			\rule[-1ex]{0pt}{3.5ex} type & 
			String identifying the name of the procedure class to instantiate. \\
			\hline
			
			\rule[-1ex]{0pt}{3.5ex} hw & 
			
			\begin{tabular}{l}
				String name of the instantiated instrument drivers that the procedure needs access to.
			\end{tabular} \\
			\hline
			
			\rule[-1ex]{0pt}{3.5ex} records & 
			Dictionary to specify output data products and their associated viewer/recorder queues. \\
			\hline
			
			\rule[-1ex]{0pt}{3.5ex} kwargs & 
			Dictionary with procedure initialization arguments. \\
			\hline			
		\end{tabular}
	\end{center}
\end{table}

\begin{table}[H]
	\caption{SLT \textbf{hardware configuration dictionary} format. The \textbf{2) INSTRUMENT\_SUITE} variable is a python list of dictionaries with these key/value pairs.} 
	\label{table:HardwareConfig}
	\begin{center}       
		\begin{tabular}{|l|l|} 
			\hline
			\rule[-1ex]{0pt}{3.5ex}  \textbf{Key} & \textbf{Value Description}  \\
			\hline
			
			\rule[-1ex]{0pt}{3.5ex} instance\_name & 
			\begin{tabular}{l}
			String to provide a name for the instrument driver instance.
			\end{tabular} \\
			\hline
			
			\rule[-1ex]{0pt}{3.5ex} resource\_name & 
			String identifying the instrument communication port. \\
			\hline
	
			\rule[-1ex]{0pt}{3.5ex} manufacturer & 
			
			\begin{tabular}{l}
			String name of the manufacturer of the instrument as labeled in \\ the SLT/PyMeasure instrument repository
			\end{tabular} \\
			\hline

			\rule[-1ex]{0pt}{3.5ex} instrument & 
			String name of the instrument driver class. \\
			\hline
			
			\rule[-1ex]{0pt}{3.5ex} params & 
			Dictionary with instrument parameters to set upon initialization. \\
			\hline
			
			\rule[-1ex]{0pt}{3.5ex} kwargs & 
			Dictionary with arguments for the driver initialization method. \\
			\hline		
		\end{tabular}
	\end{center}
\end{table} 

\begin{table}[H]
	\caption{SLT \textbf{viewer configuration dictionary} format. The \textbf{3) VIEWERS} variable is a python list of dictionaries with these key/value pairs.} 
	\label{table:ViewConfig}
	\begin{center}       
		\begin{tabular}{|l|l|} 
			\hline
			\rule[-1ex]{0pt}{3.5ex}  \textbf{Key} & \textbf{Value Description}  \\
			\hline
			
			\rule[-1ex]{0pt}{3.5ex} instance\_name & 
			\begin{tabular}{l}
				String name of the viewer instance.
			\end{tabular} \\
			\hline
			
			\rule[-1ex]{0pt}{3.5ex} type & 
			\begin{tabular}{l}
				String type of the viewer. Currently "LineViewer" and "ImageViewer" are supported \\ to display time stream and image data respectively. \\
				
			\end{tabular} \\			
			\hline
			
			\rule[-1ex]{0pt}{3.5ex} kwargs & 
			Dictionary with viewer initialization arguments. \\
			\hline			
		\end{tabular}
	\end{center}
\end{table}

\begin{table}[H]
	\caption{SLT \textbf{recorder configuration dictionary} format. The \textbf{4) RECORDERS} variable is a python list of dictionaries with these key/value pairs.} 
	\label{table:RecConfig}
	\begin{center}       
		\begin{tabular}{|l|l|} 
			\hline
			\rule[-1ex]{0pt}{3.5ex}  \textbf{Key} & \textbf{Value Description}  \\
			\hline
			
			\rule[-1ex]{0pt}{3.5ex} instance\_name & 
			\begin{tabular}{l}
				String name of the recorder instance.
			\end{tabular} \\
			\hline
			
			\rule[-1ex]{0pt}{3.5ex} type & 
			String type of the recorder to specify the output file format.\\
			\hline
			
			\rule[-1ex]{0pt}{3.5ex} kwargs & 
			Dictionary with recorder initialization arguments. \\
			\hline			
		\end{tabular}
	\end{center}
\end{table}

\begin{table}[H]
	\caption{SLT \textbf{controller configuration dictionary} format. The \textbf{5) CONTROLLERS} variable is a python list of dictionaries with these key/value pairs.} 
	\label{table:CntrlConfig}
	\begin{center}       
		\begin{tabular}{|l|l|} 
			\hline
			\rule[-1ex]{0pt}{3.5ex}  \textbf{Key} & \textbf{Value Description}  \\
			\hline
			
			\rule[-1ex]{0pt}{3.5ex} instance\_name & 
			\begin{tabular}{l}
			String name of the controller instance.
			\end{tabular} \\
			\hline
			
			\rule[-1ex]{0pt}{3.5ex} type & 
			String type of the controller. Either "InstrumentController" or "ProcedureController"\\
			\hline
			
			\rule[-1ex]{0pt}{3.5ex} hw & 
			\begin{tabular}{l}
			String name of the hardware to be controlled if the type is "InstrumentController."
			\end{tabular} \\
			\hline
			
			\rule[-1ex]{0pt}{3.5ex} procedure & 
			\begin{tabular}{l}
				String name of the procedure to be controlled if the type is "ProcedureController."
			\end{tabular} \\
			\hline

			\rule[-1ex]{0pt}{3.5ex} kwargs & 
			Dictionary with controller initialization arguments. \\
			\hline			
		\end{tabular}
	\end{center}
\end{table}

\subsection{Class Descriptions and User Interface} \label{sect:ClassDescription_UI}

\subsubsection{Procedures}

\label{sect:Procedures}

\begin{figure}[h]
	\centering
	\includegraphics[width=\linewidth]{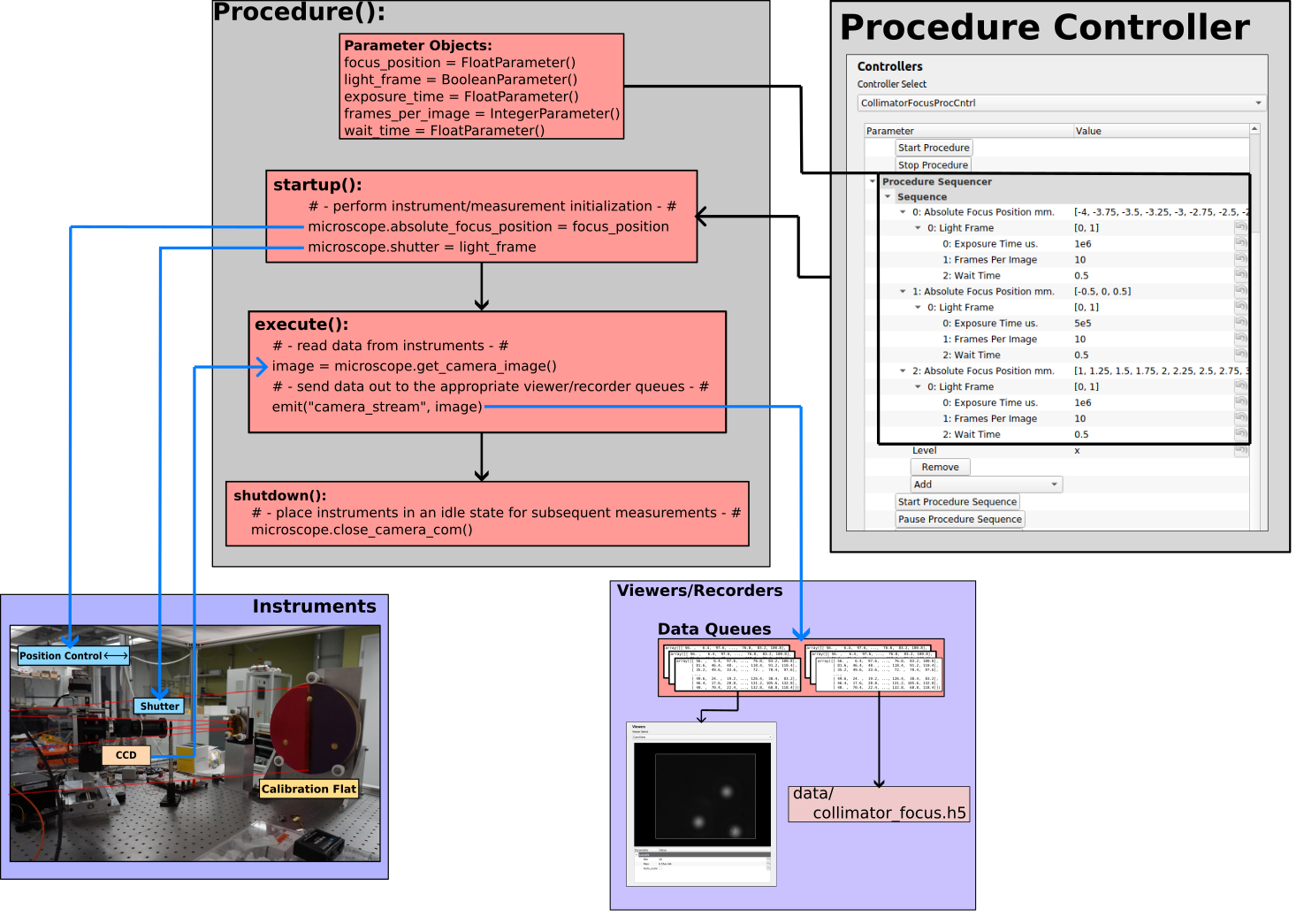}
	\caption{SLT Procedure Execution and Structure: Procedures in SLT and PyMeasure operate with the same basic principle, but with different mechanisms for how output data is handled. Shown here is a basic example procedure for the collimator calibration measurement (sect. \ref{section:Focus Cal}). In this procedure, the startup() method moves the microscope into position and opens or closes the shutter. In execute() images are recorded from the ccd camera and the emit() method is used to send images out to the "camera\_stream" recorder and viewer queues. A Procedure Controller is generated to allow measurement operators to configure the microscope position, shutter state, and exposure time.}
	\label{fig:ProcedureExecution}
\end{figure}

The central element of both the PyMeasure and SLT packages is the \textbf{Procedure} class. As mentioned in Figure \ref{fig:SLT_Architecture}, \textbf{Procedures} are used to define the logical flow of a measurement. Procedures contain the scripts which interact directly with instruments, collect and process data, and send data to Viewers and Recorders. The basic working principle of a procedure in PyMeasure and SLT is the same with SLT modifying the manner in which procedure output data is handled. Procedures in PyMeasure and SLT contain the following components which are illustrated in the flowchart of Figure \ref{fig:ProcedureExecution}:

\begin{enumerate}
	\item \textbf{Parameter Objects:} Classes defining the configurable elements of a Procedure. Parameter objects map directly to the graphical elements found in \textit{Procedure Controllers} (sect. \ref{sect:Control}) so that they can be set and modified via a graphical interface.
	\item \textbf{startup():} The initial method which executes first when a Procedure is started. In general, startup() is used to perform the initial configuration of instrumentation in a testbed before the real measurement begins.
	\item \textbf{execute():} This method always contains the main body of the measurement control code and executes after startup(). This is where data is collected, processed, and sent out for display and archival.
	\item \textbf{shutdown():} The final method executing after execute(). Here, all instrumentation is placed in an idle state before the next measurement is run. \\
	\item \textbf{emit():} This method is used to send data out for display and archival. In PyMeasure, Python's built-in \textbf{logging} package is used to write data records sent out by emit() to csv log files. Since the logging package only supports plain-text data, the output file formats supported by PyMeasure are inherently limited. In SLT we modify this mechanism to support multiple output file formats and to separate data visualization from data archival. As illustrated in Figure \ref{fig:SLT_Architecture}, all Viewers and Recorders have associated \textbf{queue} objects. When emit() is called in SLT, two arguments are provided: The first is a string that identifies which queue objects data should be sent to and the second is the data itself. Copies of the data are then generated and placed onto every desired queue object. When the Viewer and/or Recorder associated with a given queue object detects that new data is present, the data is retrieved from the queue and the appropriate display/archival task is performed. This approach has the advantage that python queue objects have no type preference so any data type can be input to the queue. Accordingly, data of all shapes and sizes can be supported so long as the appropriate Viewer/Recorder class to handle the data has been implemented. In addition, queues in SLT are independent between each instance of a Viewer or Recorder. This allows information to be displayed in a Viewer but not necessarily saved out with a Recorder and vice/versa. Or, multiple different Viewers and Recorders can be configured to process the same output data. An argument against this approach is that each time emit() is called, copies of the data object must be created for every queue so for large data objects, CPU memory could become a limiting factor. However, with modern CPU memory sizes this limitation effectively does not exist for the vast majority of practical SLT applications (certainly all applications in SPHEREx testing).
\end{enumerate}

\subsubsection{Instruments} \label{sect:Instruments}

SLT adopts the same instrument driver framework as PyMeasure. This allows seamless integration of the extensive repository of existing PyMeasure drivers into SLT. Additionally, users of SLT can develop drivers for instruments not already supported by SLT or PyMeasure by following the detailed instructions in the \emph{Adding Instruments} section of the PyMeasure documentation \cite{PyMeasure}.

\subsubsection{Viewers and Recorders} \label{sect:View_Rec}

All viewers and recorders sit idle until data is placed on their associated queues, upon which they retrieve the data object and perform their archival or display task. To date, Viewer classes have been developed for multi-channel timestream and 2-dimensional image data display. Recorder classes to support data output to CSV, HDF5, and MAT file formats as well as Structured Query Language (SQL) databases exist.

\subsubsection{Controllers} \label{sect:Control}

SLT implements two main types of controllers: \textbf{Instrument Controllers} and \textbf{Procedure Controllers}. Instrument Controllers couple directly to instrument drivers and allow manual control of any instrument supported by SLT and PyMeasure through graphical interfaces. Procedure Controllers are used to execute individual procedures as well as several procedures in a sequence. Users can set and modify the \textbf{Parameter Objects} for a given procedure and create \textit{nested for loops} where the parameters of the procedure are modified at each step of the loop.

\subsubsection{User Interface} \label{sect:Interface}

\begin{figure}[H]
	\centering
	\includegraphics[width=\linewidth]{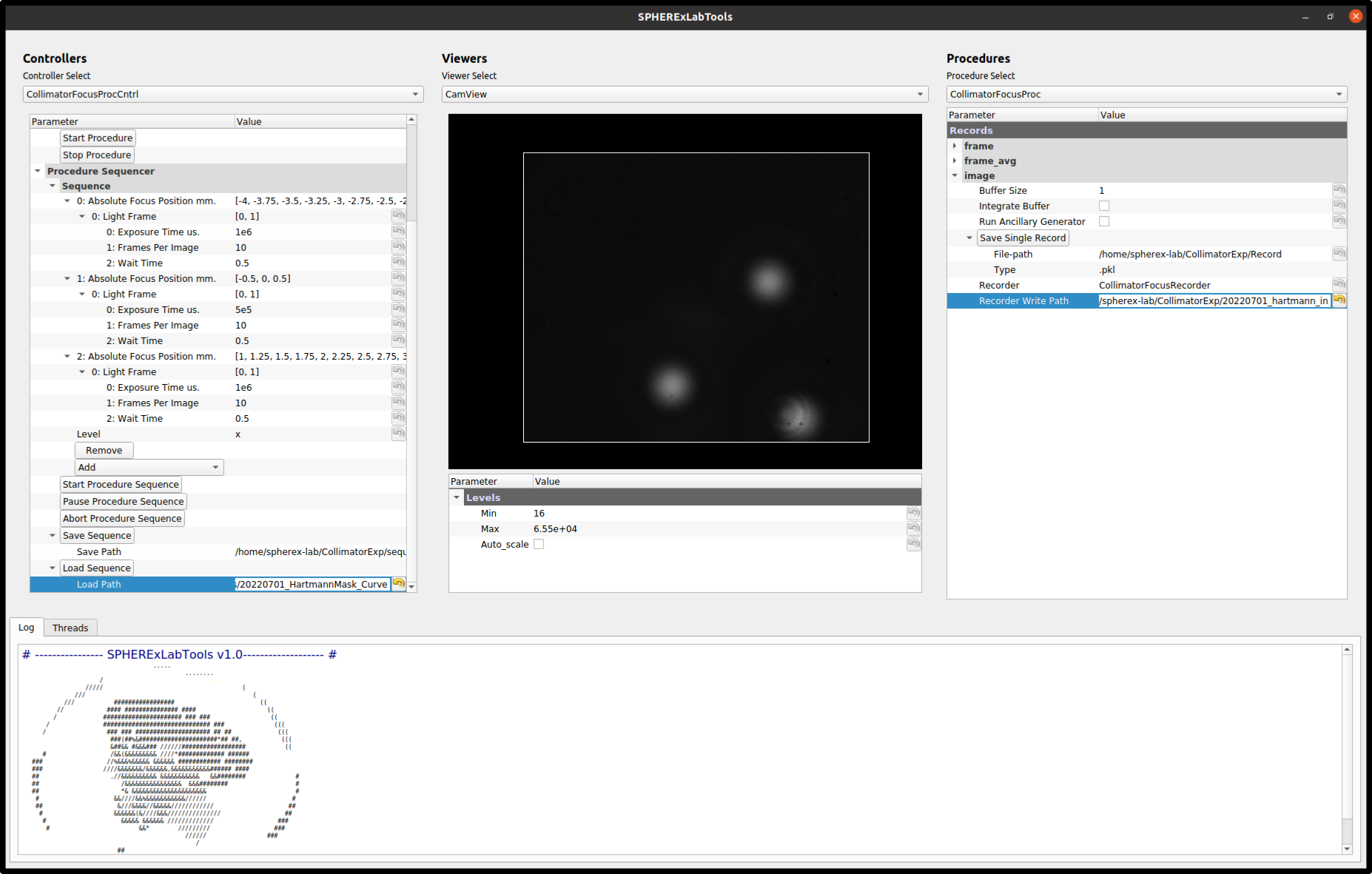}
	\caption{SLT interface: \textbf{controllers} are embedded in the left panel, \textbf{viewers} in the center, while \textbf{procedures} and their data products are displayed in the right panel. \textbf{Log messages and errors} occurring throughout a measurement are displayed in the bottom panel. Drop-down menus in each panel allow users to switch between the various controllers, viewers, and procedures.} 
	\label{fig:SltInterface}
\end{figure}

Shown in Figure \ref{fig:SltInterface} is the SLT user interface generated with the example \textit{Experiment Control Package} described in section \ref{section:Focus Cal}. Note that all graphical elements of SLT are implemented using the PyQtGraph graphics and gui library \cite{PyQtGraph}.

\section{Example: SLT for spherex spectral calibration} \label{section:Spectral Cal}

\begin{figure}[H]
    \centering
    \includegraphics[width=0.7\linewidth]{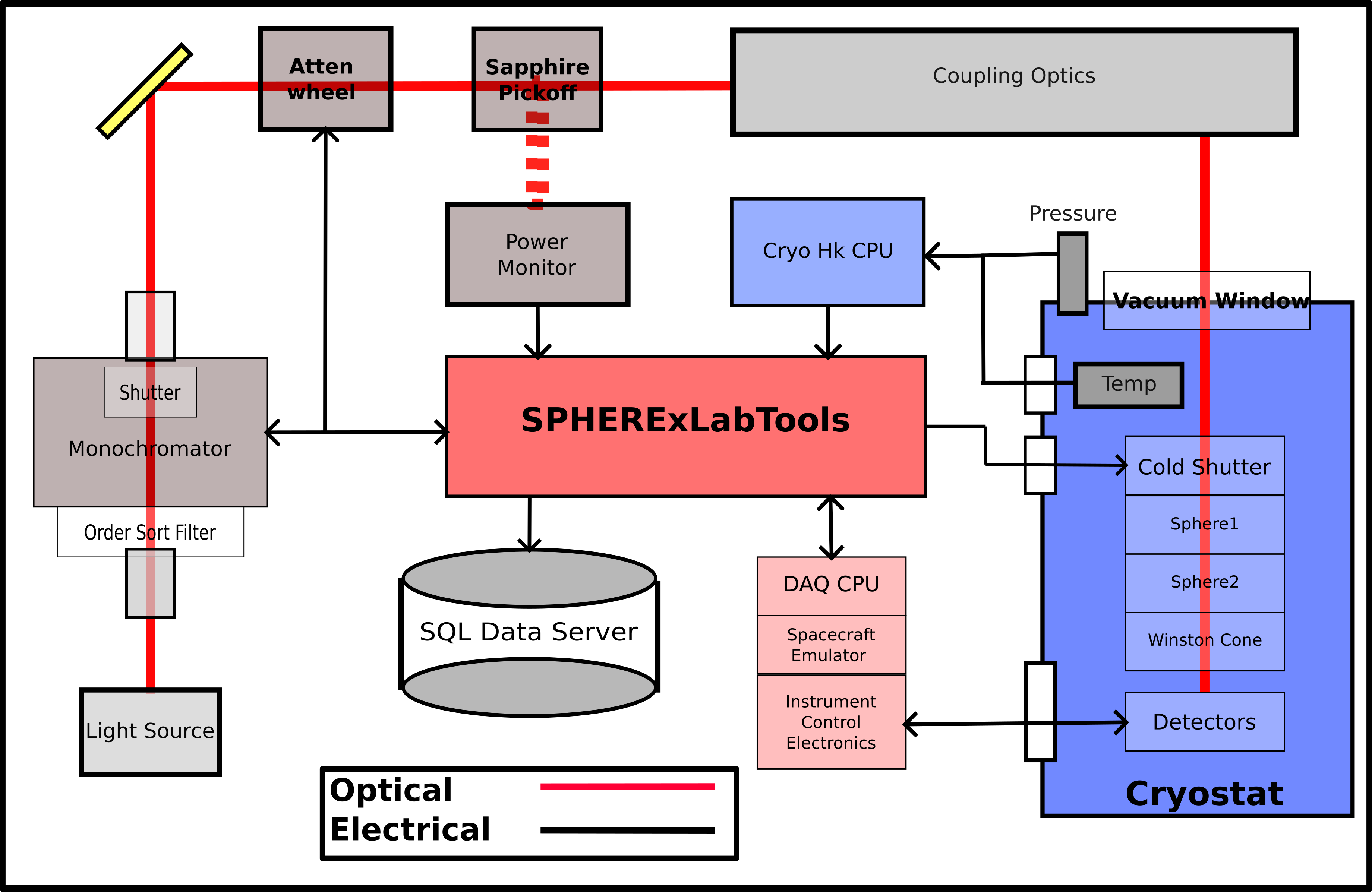}
    \caption{SPHEREx Spectral Calibration Measurement Schematic: Here, SLT drives the monochromator, a set of shutters including a cold shutter inside of the cryostat, and a neutral density filter attenuation wheel. SLT also communicates directly with the detector data acquisition system while logging the cryostat input optical power level and pressure/temperature housekeeping data. SLT uploads all metadata, housekeeping, and detector data to a central SQL server.}
    \label{fig:SpectralCal}
\end{figure}

\begin{figure}[H]
	\centering
	\includegraphics[width=0.7\linewidth]{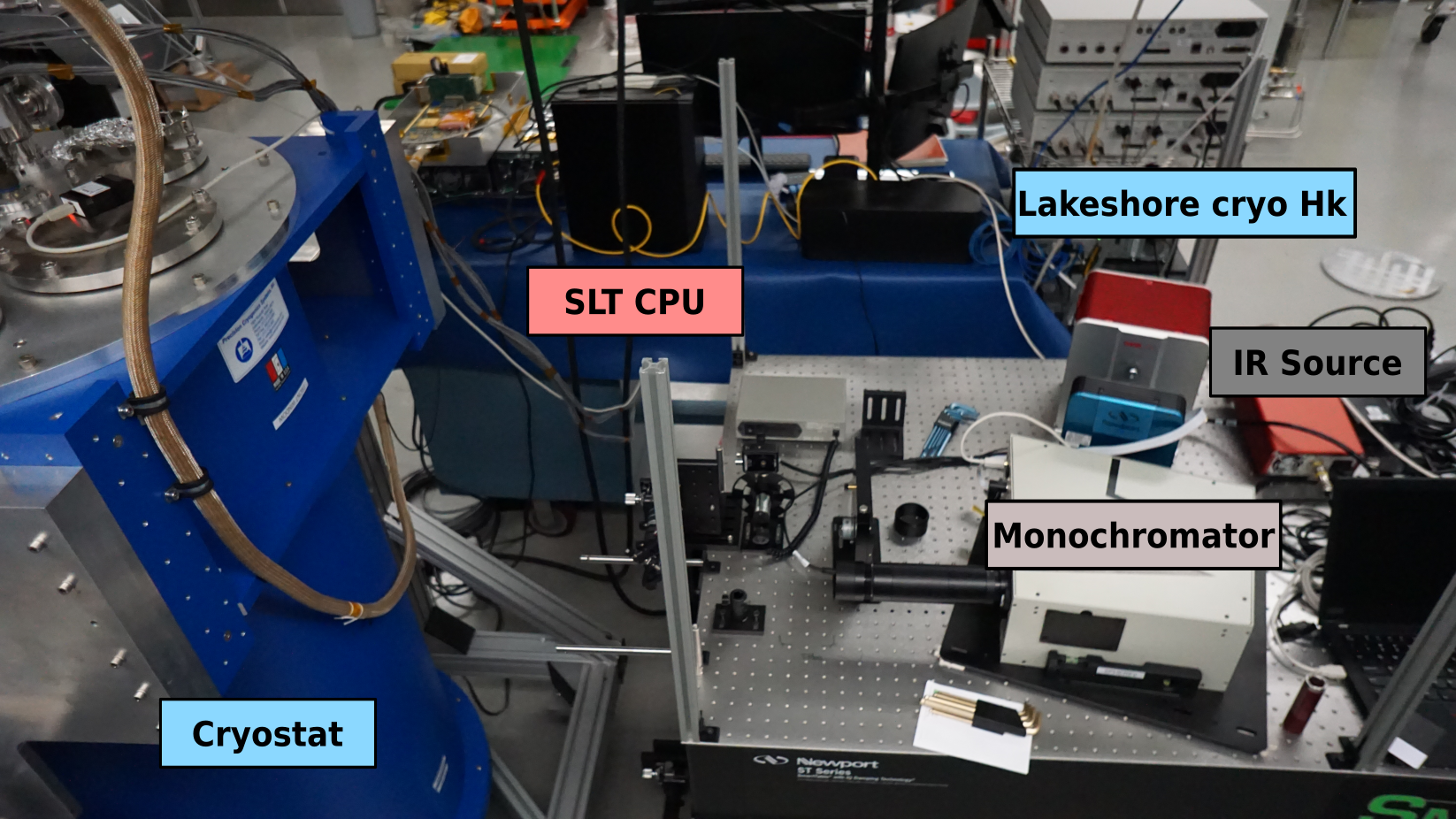}
	\caption{Prototype focal plane array spectral calibration testbed}
	\label{fig:SpecCalBBTestBed}
\end{figure}

\begin{figure}[H]
	\centering
	\includegraphics[width=0.65\linewidth]{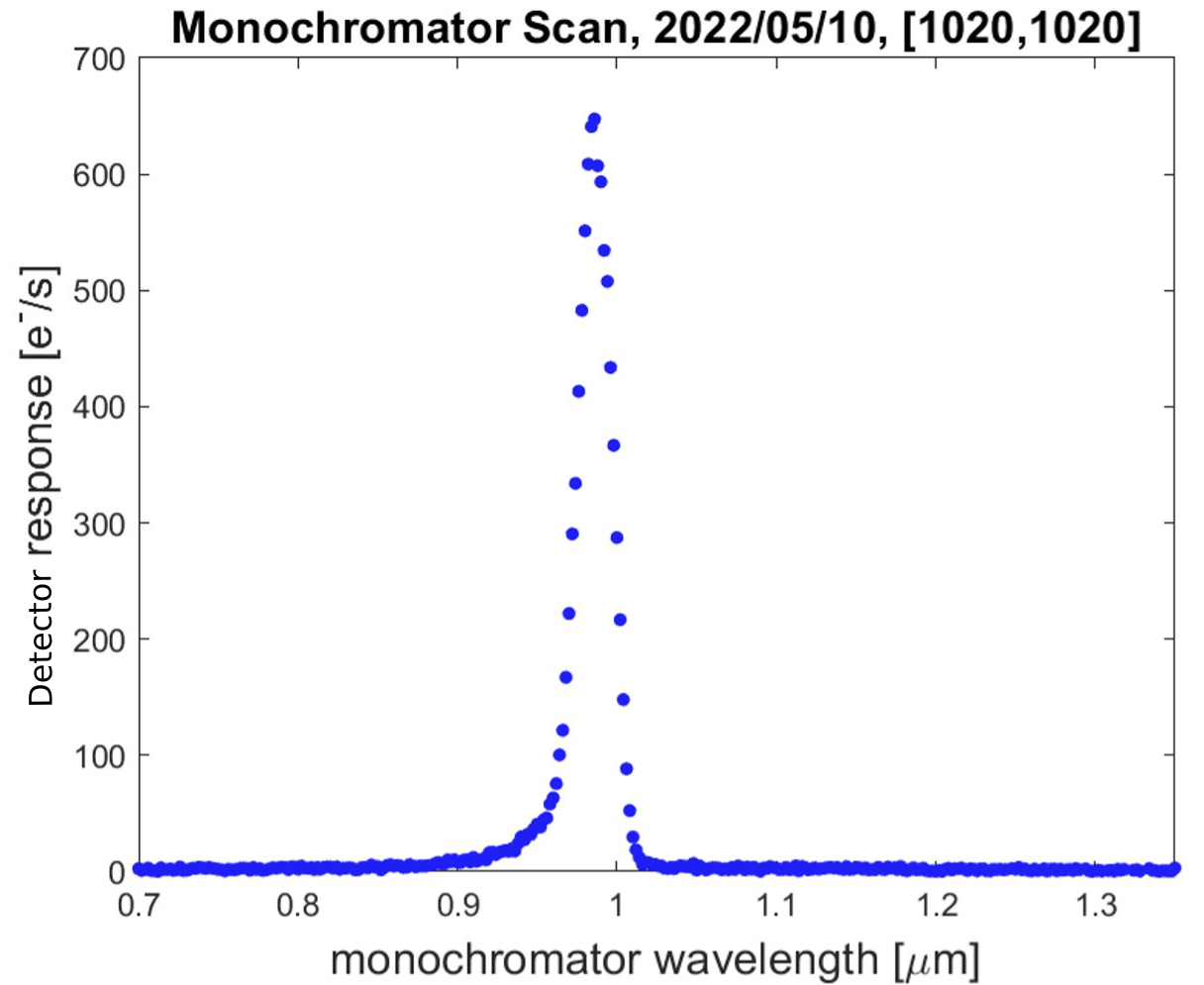}
	\caption{Spectral calibration bandpass results for a single pixel of the prototype focal plane assembly. The spectral information of this sample pixel is obtained by scanning the monochromator from 0.7 to 1.3 um at 2 nm wavelength steps. This 10 hour measurement procedure is performed in a fully automated manner by SLT, running the \textit{SpecCalProc} procedure described in Table \ref{table:SpecCalConfig}.}
	\label{fig:SpecCalBandpass}
\end{figure}

The SPHEREx spectral calibration measurement aims to deliver per-pixel spectral bandpass response profiles for every pixel in the SPHEREx detectors. This measurement is important as precise knowledge of the SPHEREx spectral response is required to accurately measure galaxy redshifts and ice absorption features in the Milky Way. We accomplish SPHEREx spectral calibration by coupling a broadband light source to a monochromator whose output beam is collimated and input to the SPHEREx optics. The monochromator contains diffraction gratings whose diffraction pattern is reflected through a small slit such that a monochromatic light source of a desired wavelength is generated. The collimated, monochromatic light is then scanned across the entire SPHEREx band in less than 1 nm wavelength steps. SPHEREx detector exposures are recorded at each wavelength. Figure \ref{fig:SpectralCal} provides a block diagram of the measurement setup and Figure \ref{fig:SpecCalBBTestBed} shows the testbed for the spectral calibration measurement of a prototype focal plane assembly model. Table \ref{table:SpecCalConfig} describes the SLT \textit{experiment control package} configuration variables for the measurement while bandpass results for the prototype calibration are shown for a single pixel in Figure \ref{fig:SpecCalBandpass}. 

\begin{table}[H]
	\caption{SLT experiment control package for spectral calibration} 
	\label{table:SpecCalConfig}
	\begin{center}       
		\begin{tabular}{|l|l|} 
			\hline
			\rule[-1ex]{0pt}{3.5ex}  \textbf{Variable} & \textbf{Description}  \\
			\hline
			\rule[-1ex]{0pt}{3.5ex} 1) PROCEDURES & 
			\begin{tabular}{l}
				\textbf{\textit{PowerLogProc:}} Continuously logs the input optical power level. \\
				\textbf{\textit{SpecCalProc:}} Sets the monochromator wavelength, attenuation wheel \\
									  position, and communicates with the detector data acquisition system. \\
									  Calculates the average cryostat temperature and pressure. Sends all \\
									  instrument metadata, cryo housekeeping, and detector exposures to the \\ \textit{SpecCalSql} recorder.
			\end{tabular} \\
			\hline
			\rule[-1ex]{0pt}{3.5ex} 2) INSTRUMENT\_SUITE & 
			\begin{tabular}{l}
				Contains hardware configuration dictionaries for the monochromator, \\ attenuation wheel, power monitor detector readout, cryostat temperature \\ and pressure readout, and the detector data acquisition system.
			\end{tabular} \\
			\hline
			\rule[-1ex]{0pt}{3.5ex} 3) VIEWERS & 
			\begin{tabular}{l}
				\textbf{\textit{PowerView:}} Displays the time-stream input optical power level data \\ generated by the \textit{PowerLogProc} procedure.
			\end{tabular} \\
			\hline
			\rule[-1ex]{0pt}{3.5ex} 4) RECORDERS & 
			\begin{tabular}{l}
				\textbf{\textit{SpecCalSql:}} Writes data from \textit{SpecCalProc} to a SQL database.
			\end{tabular} \\
			\hline
			\rule[-1ex]{0pt}{3.5ex} 5) CONTROLLERS & 
			\begin{tabular}{l}
				Configuration dictionaries to generate manual instrument controllers for all \\ instruments in INSTRUMENT\_SUITE and procedure controllers for all procedures \\ in PROCEDURES.
			\end{tabular} \\
			\hline
		\end{tabular}
	\end{center}
\end{table}

\section{Example: SLT for spherex focus calibration} \label{section:Focus Cal}

The SPHEREx focus calibration quantifies telescope focus error by determining the axial displacement of the focal plane assembly hardware from the true focal plane of the telescope. Focus calibration is important to verify that the SPHEREx PSF is such that a sufficient fraction of background sky pixels and source masking can be achieved to produce a spatial power spectrum measurement of the extragalactic background light. We perform this calibration through two measurements: collimator calibration and focus error. The collimator calibration allows us to position a broadband pinhole light-source relative to an off-axis-paraboloid (OAP) such that we generate a collimated beam. We then steer the collimated beam into the SPHEREx optics and move the pinhole through a set of positions so that the beam deviates from collimation. We record SPHEREx exposures at each pinhole position and find the position at which the SPHEREx PSF is minimized. The difference between the initial pinhole position (at which the beam is collimated) and the final pinhole position (at which the SPHEREx PSF is a minimum) determines SPHEREx focus error. We perform the collimator calibration, i.e. positioning of a pinhole to generate a collimated beam, through the following procedure:

\begin{enumerate}
	\item An LED is fiber coupled to a microscope mounted on a motorized stage. 
    \item A flat is placed inside the cryo-chamber in front of the SPHEREx optics to reflect incoming optical signal. 
    \item The LED source propagates through the collimating optics whose output beam is steered into the flat and reflected back into the collimating optics. 
    \item The reflected beam is re-imaged onto the surface of a CCD camera in the microscope and the spot size on the camera is calculated. 
    \item The microscope is scanned through several positions at which a camera spot size is calculated such that a microscope position vs. spot size curve is generated. 
    \item The minimum of a fit to the spot size vs. microscope position curve is taken to be the "best focus position" of the collimator. This is the position at which the pinhole source for the focus error measurement should be placed to generate a collimated beam.
\end{enumerate}

\begin{figure}[H]
    \centering
    \includegraphics[width=0.7\linewidth]{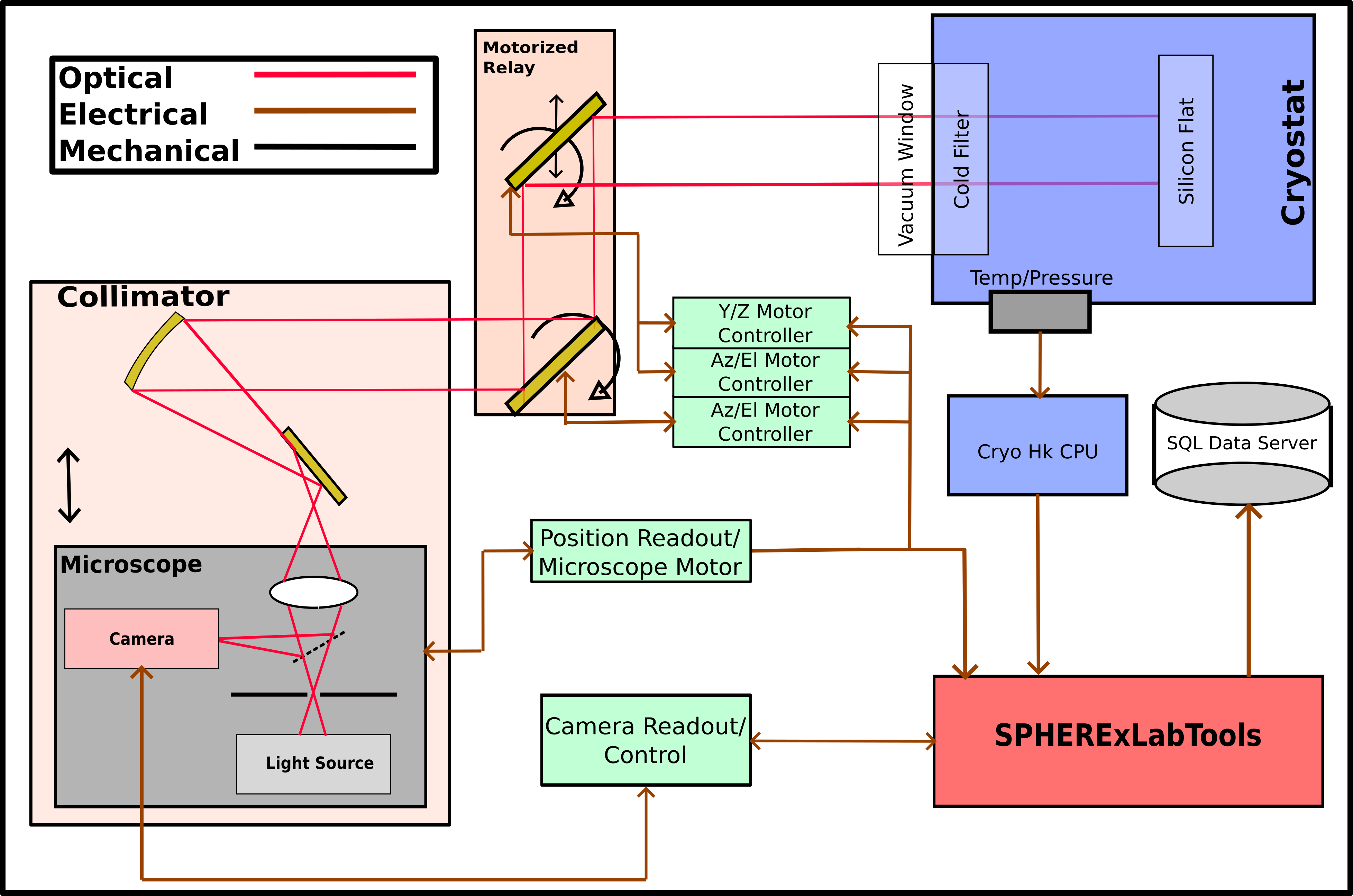}
    \label{fig:FocusCal_CollimatorCalibration}
    \caption{\textbf{Collimator calibration measurement schematic.} Here, SLT drives the motorized microscope position and reads out this position. SLT drives a motorized relay to steer the beam into the cryo chamber, and implements a module for CCD camera readout and basic real-time processing. Cryo housekeeping data is logged and written to a SQL data server along with all position information and CCD camera images. Collimator focus curves can be generated on the fly to facilitate a rapid transition to the focus error measurement.}
\end{figure}

\begin{figure}[H]
    \centering
    \includegraphics[width=0.6\linewidth]{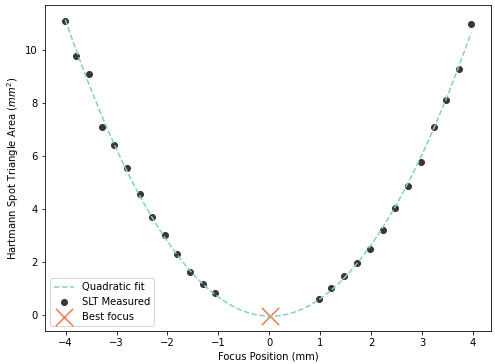}
    \caption{Collimator calibration focus curve: Black dots represent calculated beam sizes, the blue dash is the quadratic fit to the beam size data, and the red X is the minimum of the fit which is taken to be the "best focus position" of the collimator.}
    \label{fig:Fcurve}
\end{figure}

\begin{figure}[H]
	\centering
	\includegraphics[width=0.7\linewidth]{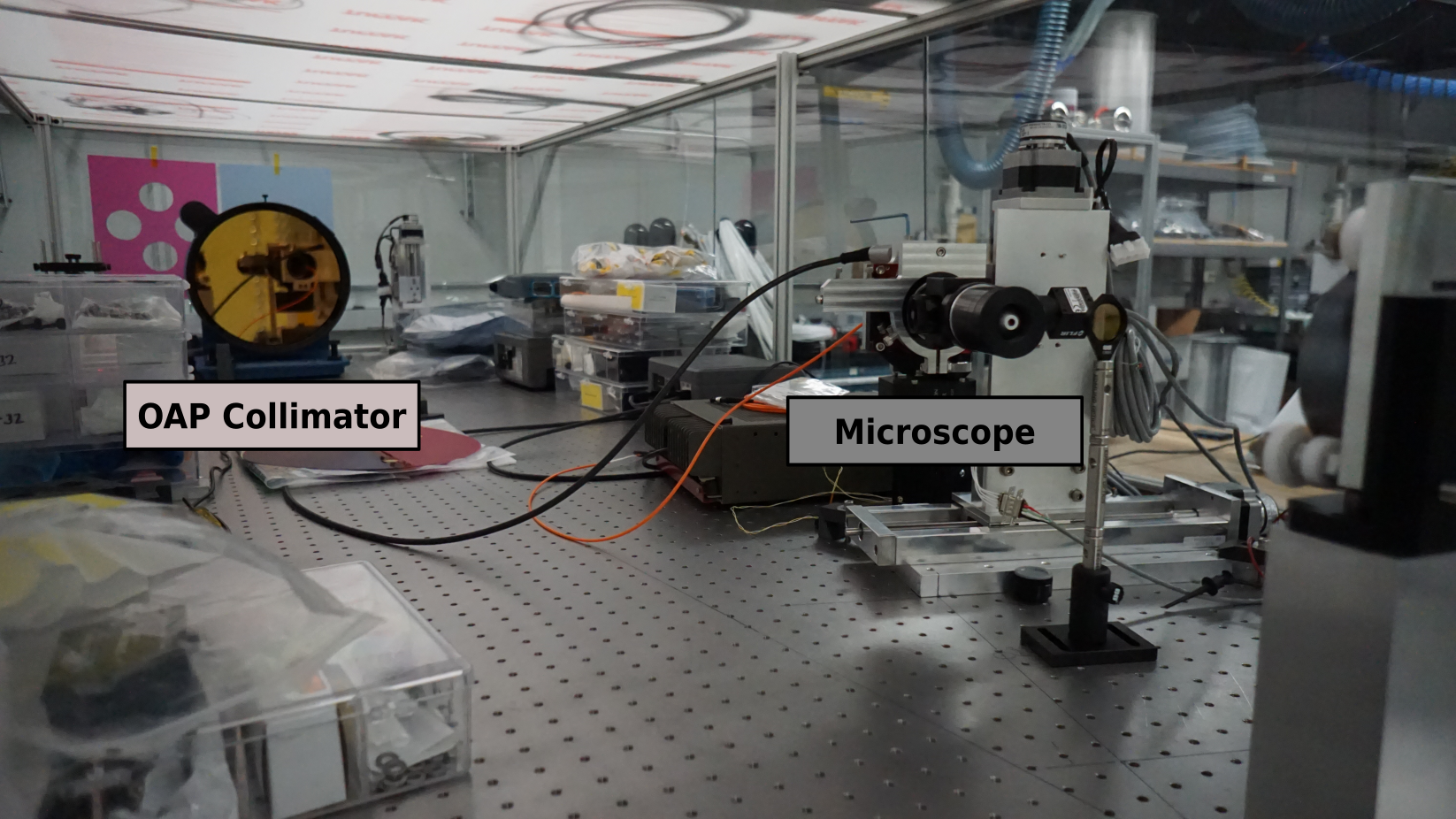}
	\caption{Microscope used in the collimator calibration measurement and off-axis paraboloid collimating mirror.}
	\label{fig:Collimator}
\end{figure}

\begin{figure}[H]
	\centering
	\includegraphics[width=0.7\linewidth]{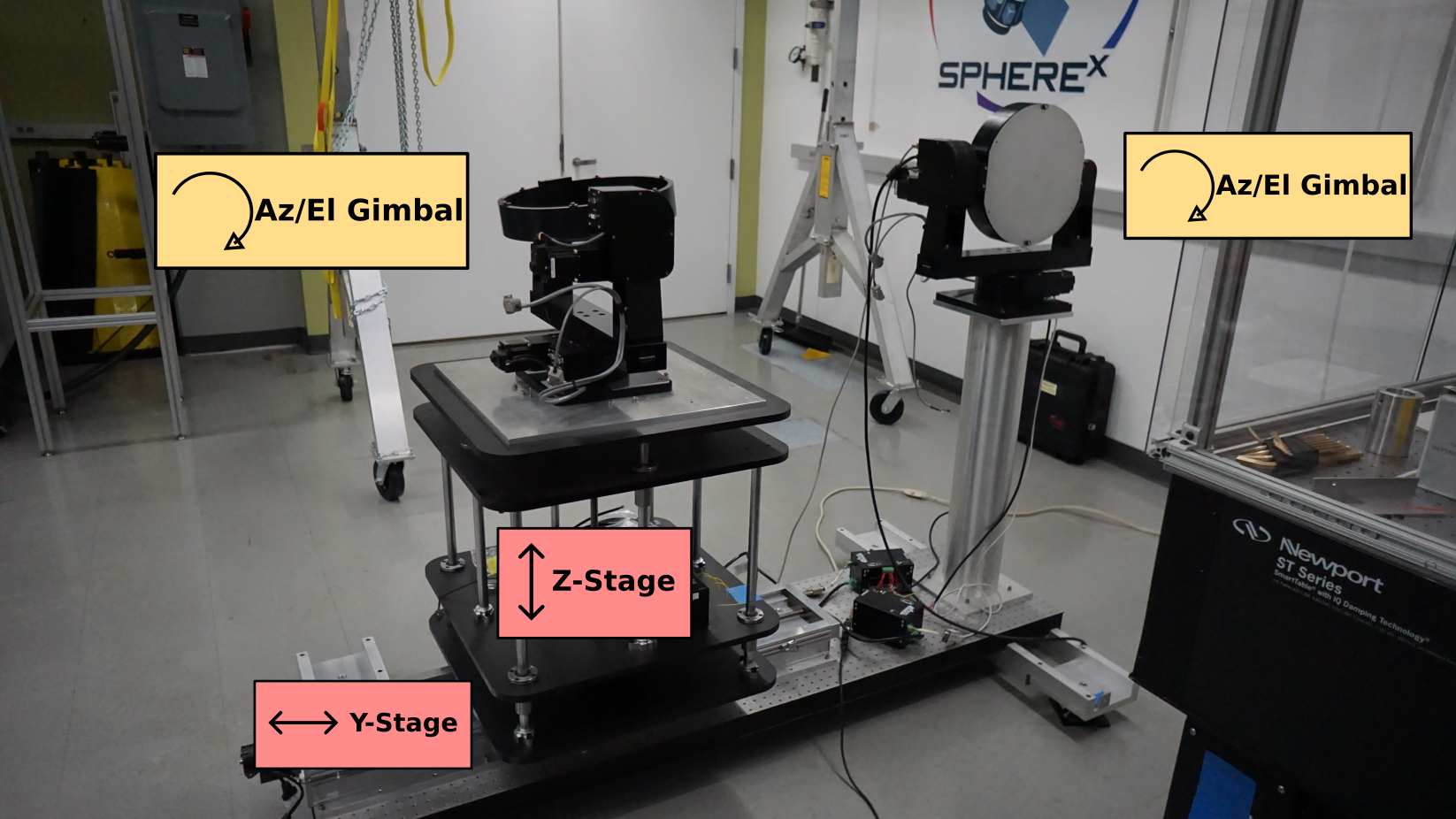}
	\caption{Motorized relay used to steer the beam into the SPHEREx optics to explore the whole field-of-view during focus calibration.}
	\label{fig:Relay}
\end{figure}

The \textit{SLT experiment control package} utilized for the collimator calibration measurement is described in Table \ref{table:CollimatorCalConfig} and an example collimator calibration focus curve generated by SLT is shown in Figure \ref{fig:Fcurve}. The collimating optics / microscope and motorized relay setup are pictured in Figures \ref{fig:Collimator} and \ref{fig:Relay} respectively.

\begin{table}[H]
	\caption{SLT experiment control package for collimator calibration} 
	\label{table:CollimatorCalConfig}
	\begin{center}       
		\begin{tabular}{|l|l|} 
			\hline
			\rule[-1ex]{0pt}{3.5ex}  \textbf{Variable} & \textbf{Description}  \\
			\hline
			\rule[-1ex]{0pt}{3.5ex} 1) PROCEDURES & 
			\begin{tabular}{l}
				\textbf{\textit{CamViewProc:}} Continuously records frames from the CCD to send out \\
				 to the \textit{CamView} image viewer. \\
				\textbf{\textit{CollimatorFocusProc:}} Drives the microscope and relay motors. Reads positions \\
				and CCD camera images. Records average cryo temp/pressure. Sends images to \\
				\textit{CamView} and images + position information to the \textit{CollimatorFocusRecorder}. \\
			\end{tabular} \\
			\hline
			\rule[-1ex]{0pt}{3.5ex} 2) INSTRUMENT\_SUITE & 
			\begin{tabular}{l}
				Contains hardware configuration dictionaries for the CCD camera, the microscope \\
				drive motor and microscope position readout, the motorized relay drive motors, \\
				cryostat temperature and pressure readout, and the detector data acquisition system.
			\end{tabular} \\
			\hline
			\rule[-1ex]{0pt}{3.5ex} 3) VIEWERS & 
			\begin{tabular}{l}
				\textbf{\textit{CamView:}} View live camera images recorded from \textit{CamViewProc} or \\
				\textit{CollimatorFocusProc}
			\end{tabular} \\
			\hline
			\rule[-1ex]{0pt}{3.5ex} 4) RECORDERS & 
			\begin{tabular}{l}
				\textbf{\textit{CollimatorFocusRecorder:}} Write images + position information from \\
				\textit{CollimatorFocusProc} to an HDF5 file.
			\end{tabular} \\
			\hline
			\rule[-1ex]{0pt}{3.5ex} 5) CONTROLLERS & 
			\begin{tabular}{l}
				Configuration dictionaries to generate manual instrument controllers for all \\ instruments in INSTRUMENT\_SUITE and procedure controllers for all procedures \\ in PROCEDURES.
			\end{tabular} \\
			\hline
		\end{tabular}
	\end{center}
\end{table}

\section{conclusion} \label{section:Conclusion}

We have presented a high-level overview of the SPHERExLabTools (SLT) data acquisition and instrument control package utilized in SPHEREx optics integration and testing. SLT builds upon and adapts the existing open source packages, PyMeasure \cite{PyMeasure} and PyHK \cite{PyHk} and presents a framework for a general purpose laboratory data acquisition and instrument control package. The effectiveness of SLT has been demonstrated through the delivery of early SPHEREx calibration data products, including the \textit{focal plane prototype spectral response} \ref{section:Spectral Cal} and \textit{collimator calibration} \ref{section:Focus Cal} measurements. SLT development is ongoing and the package continues to improve. An open source release along with work to integrate key features of SLT into the PyMeasure project are planned, with the long term vision of helping to establish an open-source standard in laboratory data-acquisition and instrument control.    

\acknowledgments 

This work was carried out at the Jet Propulsion Laboratory, California Institute of Technology, under a contract with the National Aeronautics and Space Administration (80NM0018D0004).

\bibliography{report} 
\bibliographystyle{spiebib} 

\end{document}